\documentclass[aps,showpacs,twocolumn]{revtex4}
\usepackage{epsfig}
\usepackage{amsmath}
\usepackage{booktabs}
\usepackage{makecell}
\begin{document}

\title{Systematical investigation on the stability of doubly heavy tetraquark states}

\author{Chengrong Deng$^{a}{\footnote{crdeng@swu.edu.cn}}$,
        Hong Chen$^a{\footnote{chenh@swu.edu.cn}}$,
       and Jialun Ping$^b{\footnote{jlping@njnu.edu.cn}}$}
\affiliation{$^a$School of Physical science and technology, Southwest University, Chongqing 400715, China}
\affiliation{$^b$School of Physical science and technology, Nanjing Normal University, Nanjing 210097, China}

\begin{abstract}
We systematically investigate the stability of the doubly heavy tetraquark states $[QQ][\bar{q}\bar{q}]$ ($Q=c$ and $b$, $q=u$, $d$ and $s$)
within the framework of the alternative color flux-tube model involving a multibody confinement potential, $\sigma$-exchange, one-gluon-exchange and one-Goldstone-boson-exchange interactions. Our numerical analysis indicates that the states $[bb][\bar{u}\bar{d}]$ with $01^+$ and $[bb][\bar{u}\bar{s}]$ with $\frac{1}{2}1^+$ are the most promising stable states against strong interactions. The states $[cc][\bar{u}\bar{d}]$
with $01^+$, $[bc][\bar{u}\bar{d}]$ with $00^+$, $01^+$, and $12^+$, and $[bb][\bar{u}\bar{d}]$ with $01^-$ and $12^+$ as stable states are also predicted in the model. The dynamical mechanism producing those stable doubly heavy tetraquark states are discussed in the model.
\end{abstract}

\maketitle

\section{Introduction}

Searching for exotic hadrons beyond conventional $q\bar{q}$-meson and $qqq$-baryon pictures is an extremely meaningful topic in hadronic physics because they may contain  more abundant low-energy strong interaction information than ordinary hadrons. A large amount of new hadron states have been observed since the BELLE collaboration's discovery of the charmonium-associated state $X(3872)$ in 2003~\cite{review}. Many of new hadron states can not be accommodated in the conventional $Q\bar{Q}$-meson framework, such as the charged $Z_c^+$ states, which must have a smallest quark component $c\bar{c}u\bar{d}$ due to carrying one unit charge. Tetraquark states $Q\bar{Q}q\bar{q}$ have therefore attracted much attention from theoretical physicists to describe the internal structure of new hadron states~\cite{review}. Most of new hadron states can be accommodated in the picture of tetraquark states just matching their experimental data with model values of $Q\bar{Q}q\bar{q}$~\cite{charmed,referee}. Even so, none of new hadron states is lower than its threshold up to now and can therefore decay into $Q\bar{Q}$ and $q\bar{q}$ via strong interactions~\cite{charmed,referee}.

The theoretical explorations on the question whether or not the doubly heavy tetraquark states $[QQ][\bar{q}\bar{q}]$ or $[\bar{Q}\bar{Q}][qq]$  can exist as stable states against breakup into two $Q\bar{q}$ mesons was pioneered in the early 1980s~\cite{potential}. From then on, a lot of attention has been payed to the states in various phenomenological methods, such as the MIT bag model~\cite{bag}, constituent quark model (CQM)~\cite{constituentmodel}, chiral perturbation theory~\cite{chiralperturbation}, string model~\cite{string}, lattice QCD~\cite{lqcdbb}, and QCD sum rule approach~\cite{qcdsum}. A large amount of researches indicated that the state $[bb][\bar{u}\bar{d}]$ with $01^+$ is stable against strong interactions in various theoretical framework. However, the state has not been determined because of a lack of experimental information about the strength of the interaction between two heavy quarks. The discovery of the doubly charmed baryon $\Xi_{cc}$ by the LHCb Collaboration at CERN has provided the crucial experimental input which allows this issue to be finally resolved~\cite{xicc}. Subsequently, the enthusiasms of the theoretical investigation on the doubly heavy tetraquark states $[QQ][\bar{q}\bar{q}]$ are stimulated again to search for possible stable tetraquark states~\cite{karliner,Eichten,Francis,bicudo, cqmpark,unstable,heavy,pseudoscalar,decay,francis,richard}. Undoubtedly, the stability of the doubly heavy tetraquark states are model dependent, see Table \ref{comparison}. More investigations from the different theoretical point of view should therefore be very necessary to present a comprehensive understanding on the properties of the states, which must be benefit to the future experimental searches for stable doubly heavy tetraquark states.

Recently, we have developed an alternative color flux-tube model (ACFTM) based on the lattice QCD picture and the traditional quark models~\cite{lqcd,tcqm}. The most salient feature of this model is a multibody confinement potential instead of a two-body one proportional to the color charge in the traditional quark models~\cite{ACFTM}. We systematically investigate the hidden charmed states observed in recent years within the framework of the ACFTM. It can be found that many of hidden charmed states as compact tetraquark states can be accommodated in the ACFTM, especially the charged tetraquarks $Z^+_c$. The multibody color flux-tube dynamical mechanism  is seem to be propitious to describe multiquark states from the phenomenological point of view~\cite{charmed}. We have therefore a great ambition to research the properties of the doubly heavy tetraquark states under the hypothesis of diquark-antidiquark picture within the framework of ACFTM. We concentrate more on the mass spectrum of doubly heavy tetraqurk states than on their decay properties and we investigate the dynamical mechanism bunching quarks together and affecting their binding energy. This work is attempted to broaden the theoretical horizon in the properties of the doubly heavy tetraquark states and may provide some valuable clues to the experimental establishment of the tetraquark states in the future.

This paper is organized as follows. After the introduction section, the introduction of the ACFTM is given in Sec. II. The construction of the wavefunctions of the doubly heavy tetraquark states are shown in Sec. III. The numerical results and discussions of the stable doubly heavy tetraquark states are presented in Sec. IV. The last section is devoted to list a brief summary.

\section{The alternative color flux-tube model}

The underlying theory of strong interaction is quantum Chromodynamics (QCD), which has three important properties: asymptotic freedom, color confinement, approximate chiral symmetry and its spontaneous breaking. At the hadronic scale, QCD is highly non-perturbative due to the complicated infrared behavior of the non-Abelian $SU(3)$ gauge group. At present it is still impossible for us to derive the hadron spectrum analytically from the QCD Lagrangian. The QCD-inspired CQM is therefore a useful tool in obtaining physical insight for these complicated strong interaction systems. Though the connection between the models and QCD is not clearly established and there is no sound systematics to obtain corrections, the models can provide simple physical pictures, which connect the phenomenological regularities observed in the hadron data with the underlying structure.

Color confinement is a long distance behavior whose understanding continues to be a challenge in theoretical physics. In the traditional CQM, a two-body interaction $V^{C}_{ij}$ proportional to the color charges $\mathbf{\lambda}^c_i\cdot\mathbf{\lambda}^c_j$ and $r_{ij}^n$, namely
$V^{C}_{ij}=a_c\mathbf{\lambda}^c_i\cdot\mathbf{\lambda}^c_jr_{ij}^n$, where $n=1$ or 2 and $r_{ij}$ is the distance between two quarks, was introduced to phenomenologically describe the quark confinement interaction~\cite{tcqm}. The traditional models can well describe the properties of ordinary hadrons ($q^3$ and $q\bar{q}$) because their flux-tube structures are unique and trivial. The models are known to be flawed phenomenologically because it leads to power-law van der Waals forces between color-singlet hadrons~\cite{vandewaals}. The problems are related to the fact that the models do not respect local color gauge invariance~\cite{gauge}. In addition, it also leads to anticonfinement for symmetrical color structure in the multiquark system~\cite{anticonfinement}.

LQCD calculations on mesons, baryons, tetraquark, and pentaquark states revealed that there exist flux-tube structures in the hadrons~\cite{lqcd}. In the case of a given spatial configuration of multiquark states, the confinement is a multibody interaction and can be simulated by a static potential which is proportional to the minimum of the total length of color flux-tubes. A naive flux-tube model, used in the present work, based on this picture has been constructed. It takes into account multibody confinement with harmonic interaction approximation, i.e., where the length of the color flux-tube is replaced by the square of the length to simplify the numerical calculation. In this way, the Regge trajectories are missing in the ACFTM. However, this replacement is still a good approximation for low-lying states such as the states considered in this paper.
We have calculated the $b\bar{b}$ spectrum by using quadratic and linear potentials, the results show that the differences between two models are small for the low-lying states~\cite{SNChen}. There are two theoretical arguments to support this approximation: One is that the spatial separations of the quarks (lengths of the color flux-tube) in hadrons are not large, so the difference between the linear and quadratic forms is small and can be absorbed in the adjustable parameter, the stiffness. The calculations on nucleon-nucleon interactions support the argument~\cite{linear}. The second is that we are using a nonrelativistic description of the dynamics and, as was shown long ago~\cite{goldman}, an interaction energy that varies linearly with separation between fermions in a relativistic, first order differential dynamics has a wide region in which a harmonic approximation is valid for the second order (Feynman-Gell-Mann) reduction of the equations of motion.

For an ordinary meson, the quark and anti-quark are connected by a three-dimension color flux-tube. It's confinement potential in the ACFTM can be written as
\begin{eqnarray}
V^{C}_{min}(2) = kr^2,
\end{eqnarray}
where $r$ is the separation of the quark and anti-quark, $k$ is the stiffnesses of a three-dimension color flux-tube.

According to double Y-shaped color flux-tube structures of the tetraquark state $[Q_1Q_2][\bar{q}_3\bar{q}_4]$ with diquark-antiquark configuration, a four-body quadratic confinement potential instead of linear one used in the lattice QCD can be written as,
\begin{eqnarray}
V^{C}(4)&=&k\left[ (\mathbf{r}_1-\mathbf{y}_{12})^2
+(\mathbf{r}_2-\mathbf{y}_{12})^2+(\mathbf{r}_{3}-\mathbf{y}_{34})^2\right. \nonumber \\
&+&
\left.(\mathbf{r}_4-\mathbf{y}_{34})^2+\kappa_d(\mathbf{y}_{12}-\mathbf{y}_{34})^2\right],
\end{eqnarray}
where $\mathbf{r}_1$, $\mathbf{r}_2$, $\mathbf{r}_3$ and $\mathbf{r}_4$ are particle's positions. Two Y-shaped junctions $\mathbf{y}_{12}$ and $\mathbf{y}_{34}$ are variational parameters, which can be determined by taking he minimum of the confinement potential. $\kappa_d k$ is the stiffness of a $d$-dimension color flux-tube. The relative stiffness parameter $\kappa_{d}$ is equal to $\frac{C_{d}}{C_3}$~\cite{kappa}, where $C_{d}$ is the eigenvalue of the Casimir operator associated with the $SU(3)$ color representation $d$ at either end of the color flux-tube, such as $C_3=\frac{4}{3}$, $C_6=\frac{10}{3}$, and $C_8=3$.

The minimum of the confinement potential $V^C_{min}$ can be obtained by taking the variation of $V^C$ with respect to $\mathbf{y}_{12}$ and $\mathbf{y}_{34}$, and it can be expressed as
\begin{eqnarray}
V^C_{min}(4)&=& k\left(\mathbf{R}_1^2+\mathbf{R}_2^2+
\frac{\kappa_{d}}{1+\kappa_{d}}\mathbf{R}_3^2\right),
\end{eqnarray}
The canonical coordinates $\mathbf{R}_i$ have the following forms,
\begin{eqnarray}
\mathbf{R}_{1} & = &
\frac{1}{\sqrt{2}}(\mathbf{r}_1-\mathbf{r}_2),~
\mathbf{R}_{2} =  \frac{1}{\sqrt{2}}(\mathbf{r}_3-\mathbf{r}_4), \nonumber \\
\mathbf{R}_{3} & = &\frac{1}{ \sqrt{4}}(\mathbf{r}_1+\mathbf{r}_2
-\mathbf{r}_3-\mathbf{r}_4), \\
\mathbf{R}_{4} & = &\frac{1}{ \sqrt{4}}(\mathbf{r}_1+\mathbf{r}_2
+\mathbf{r}_3+\mathbf{r}_4). \nonumber
\end{eqnarray}
The use of $V^C_{min}(4)$ can be understood here as that the gluon field readjusts immediately to its minimal configuration.

The origin of the constituent quark mass is traced back to the spontaneous breaking of $SU(3)_L\otimes SU(3)_R$ chiral symmetry and consequently constituent quarks should interact through the exchange of Goldstone bosons \cite{chnqm}. Chiral symmetry breaking suggests dividing quarks into two different sectors: light quarks ($u$, $d$ and $s$) where the chiral symmetry is spontaneously broken and heavy quarks ($c$ and $b$) where the symmetry is explicitly broken. The $SU(3)_L\otimes SU(3)_R$ chiral quark model where constituent quarks interact through pseudoscalar Goldstone bosons exchange (GBE) were very successfully applied to describe the baryon spectra~\cite{su3}, nucleon-nucleon and nucleon-hyperon interactions~\cite{NNNY}. The central part of the quark-quark interaction originating from chiral symmetry breaking can be resumed as follows~\cite{chiralmeson},
\begin{eqnarray}
V_{ij}^{B} & = & V^{\pi}_{ij} \sum_{k=1}^3 \mathbf{F}_i^k
\mathbf{F}_j^k+V^{K}_{ij} \sum_{k=4}^7\mathbf{F}_i^k\mathbf{F}_j^k \nonumber\\
&+&V^{\eta}_{ij} (\mathbf{F}^8_i \mathbf{F}^8_j\cos \theta_P
-\sin \theta_P),\nonumber\\
V^{\chi}_{ij} & = &
\frac{g^2_{ch}}{4\pi}\frac{m^3_{\chi}}{12m_im_j}
\frac{\Lambda^{2}_{\chi}}{\Lambda^{2}_{\chi}-m_{\chi}^2}
\mathbf{\sigma}_{i}\cdot
\mathbf{\sigma}_{j} \\
& \times &\left( Y(m_\chi r_{ij})-
\frac{\Lambda^{3}_{\chi}}{m_{\chi}^3}Y(\Lambda_{\chi} r_{ij})
\right),\nonumber \\
V^{\sigma}_{ij} & = &-\frac{g^2_{ch}}{4\pi}
\frac{\Lambda^{2}_{\sigma}m_{\sigma}}{\Lambda^{2}_{\sigma}-m_{\sigma}^2}
\left( Y(m_\sigma r_{ij})-
\frac{\Lambda_{\sigma}}{m_{\sigma}}Y(\Lambda_{\sigma}r_{ij})
\right). \nonumber
\end{eqnarray}
where $\chi$ stands for $\pi$, $K$ and $\eta$, $Y(x)=e^{-x}/x$, $\mathbf{F}_{i,j}$ and $\mathbf{\sigma}_{i,j}$ are the flavor $SU(3)$ Gell-man matrices and spin $SU(2)$ Pauli matrices, respectively.

Besides the chiral symmetry breaking, one expects the dynamics to be governed by QCD perturbative effects, which is well known one-gluon-exchange (OGE) potential. The central part of the OGE reads~\cite{chiralmeson},
\begin{eqnarray}
V_{ij}^{G} & = & {\frac{\alpha_{s}}{4}}\mathbf{\lambda}^c_{i}
\cdot\mathbf{\lambda}_{j}^c\left({\frac{1}{r_{ij}}}-
{\frac{2\pi\delta(\mathbf{r}_{ij})\mathbf{\sigma}_{i}\cdot
\mathbf{\sigma}_{j}}{3m_im_j}}\right). \nonumber
\end{eqnarray}
where $\mathbf{\lambda}_{i,j}$ is the color $SU(3)$ Gell-man and, $\alpha_s$ is the running strong coupling constant and takes the following form~\cite{chiralmeson},
\begin{equation}
\alpha_s(\mu_{ij})=\frac{\alpha_0}{\ln\left((\mu_{ij}^{2}+\mu_0^2)/\Lambda_0^2\right)},
\end{equation}
where $\mu_{ij}$ is the reduced mass of two interacting particles. The function $\delta(\mathbf{r}_{ij})$ should be regularized~\cite{weistein},
\begin{equation}
\delta(\mathbf{r}_{ij})=\frac{1}{4\pi r_{ij}r_0^2(\mu_{ij})}e^{-r_{ij}/r_0(\mu_{ij})},
\end{equation}
where $r_0(\mu_{ij})=\hat{r}_0/\mu_{ij}$. $\Lambda_0$, $\alpha_0$, $\mu_0$ and $\hat{r}_0$ are adjustable model parameters.

The non-central parts of the OBE and OGE, tensor and spin-orbit forces, between quarks are omitted in the present calculation because, for the lowest energy states which we are interested in here, their contributions are small or zero.

To sum up, the Hamiltonian $H_n$ ($n=2$ or 4) related to the present work can be expressed as follows:
\begin{eqnarray}
H_n & = & \sum_{i=1}^n \left(m_i+\frac{\mathbf{p}_i^2}{2m_i}
\right)-T_{C}+\sum_{i>j}^4 V_{ij}+V^{C}_{min}(n), \nonumber\\
V_{ij} & = &V_{ij}^G+V_{ij}^B+V_{ij}^{\sigma}.
\end{eqnarray}
$\mathbf{p}_i$ and $m_i$ are the momentum and mass of the $i$-th quark (antiquark), respectively. $T_{c}$ is the center-of-mass kinetic energy of the states and should be deducted.

The starting point of the model study on the multiquark states is to accommodate ordinary hadrons in the model in order to determine model parameters. The mass parameters $m_{\pi}$, $m_K$ and $m_{\eta}$ in the $V^B_{ij}$ take their experimental values. The cutoff parameters $\Lambda$s and the mixing angle $\theta_{P}$ in the $V_{ij}^B$ take the values in the work~\cite{chiralmeson}. The mass parameter $m_{\sigma}$ in the interaction $V_{ij}^{\sigma}$ can be determined through the PCAC relation $m^2_{\sigma}\approx m^2_{\pi}+4m^2_{u,d}$~\cite{masssigma}. The chiral coupling constant $g_{ch}$ can be obtained from the $\pi NN$ coupling constant through
\begin{equation}
\frac{g_{ch}^2}{4\pi}=\left(\frac{3}{5}\right)^2\frac{g_{\pi NN}^2}{4\pi}
\frac{m_{u,d}^2}{m_N^2}.
\end{equation}
The values of the above fixed model parameters are given in Table \ref{fixed}. The adjustable parameters and their errors in Table \ref{adjustable} can be determined by fitting the masses of the ground states of mesons in Table \ref{mesons} using Minuit program. Once the meson masses are obtained, one can calculate the threshold of the doubly heavy tetraquark states $[QQ][\bar{q}\bar{q}]$ simply by adding the masses of two $Q\bar{q}$ mesons to identify the stability of the tetraquark states against strong interaction.

\begin{table}[ht]
\caption{Fixed model parameters.} \label{fixed}
\begin{tabular}{cccccccccccccc}
\toprule[0.8pt]
Para.        &  Valu. & Unit      &~~~&  Para.              & Valu.& Unit      &~~~& Para.                   & Vale.             & Unit      \\
$m_{ud}$     &  280   & MeV       &   &  $m_{\sigma}$       & 2.92 & fm$^{-1}$ &   & $\Lambda_{\eta}$        &  5.2              & fm$^{-1}$ \\
$m_{\pi}$    &  0.7   & fm$^{-1}$ &   &  $\Lambda_{\pi}$    & 4.2  & fm$^{-1}$ &   & $\theta_P$              & $-\frac{\pi}{12}$ & ...       \\
$m_{K}$      &  2.51  & fm$^{-1}$ &   &  $\Lambda_{\sigma}$ & 4.2  & fm$^{-1}$ &   & $\frac{g^2_{ch}}{4\pi}$ &  0.43             & ...       \\
$m_{\eta}$   &  2.77  & fm$^{-1}$ &   &  $\Lambda_{K}$      & 5.2  & fm$^{-1}$           \\
\toprule[0.8pt]
\end{tabular}
\caption{Adjustable model parameters.}\label{adjustable}
\begin{tabular}{ccccccccccc}
\toprule[0.8pt]
Para.        &   $x_i\pm\Delta x_i$  & Unit   &     Para.       &   $x_i\pm\Delta x_i$  &  Unit         \\
$m_{s}$      &   $511.78\pm0.228$    & MeV    &    $\alpha_0$   &   $4.554\pm0.018$     &   ...         \\
$m_{c}$      &   $1601.7\pm0.441$    & MeV    &    $k$          &   $217.50\pm0.230$    &  MeV$\cdot$fm$^{-2}$         \\
$m_{b}$      &   $4936.2\pm0.451$    & MeV    &    $\mu_0$      &   $0.0004\pm0.540$    &  MeV          \\
$\Lambda_0$  &   $9.173\pm0.175$     & MeV    &    $r_0$        &   $35.06\pm0.156$     &  MeV$\cdot$fm \\
\toprule[0.8pt]
\end{tabular}
\end{table}
\begin{table}[ht]
\caption{The ground state meson spectra in the three models, unit in MeV.}\label{mesons}
\begin{tabular}{ccccccc} \toprule[0.8pt]
Mesons         & ~$IJ^P$~          &     ~~~ACFTM~~~     &~Ref.~\cite{rel-meson}~ &  ~Ref.~\cite{chiralmeson}~&  ~~PDG~~      \\
\toprule[0.8pt]
$\pi$          &  $10^-$           &     $142\pm26$     &   150     &   139    &   139              \\
$K$            &  $\frac{1}{2}0^-$ &     $492\pm20$     &   470     &   496    &   496              \\
$\rho$         &  $11^-$           &     $826\pm4$      &   770     &   772    &   775              \\
$\omega$       &  $01^-$           &     $780\pm4$      &   780     &   690    &   783              \\
$K^*$          &  $\frac{1}{2}1^-$ &     $974\pm4$      &   900     &   910    &   892              \\
$\phi$         &  $01^-$           &     $1112\pm4$     &   1020    &   1020   &   1020             \\
$D^{\pm}$      &  $\frac{1}{2}0^-$ &     $1867\pm8$     &   1880    &   1883   &   1869             \\
$D^*$          &  $\frac{1}{2}1^-$ &     $2002\pm4$     &   2040    &   2010   &   2007             \\
$D_s^{\pm}$    &  $00^-$           &     $1972\pm9$     &   1980    &   1981   &   1968             \\
$D_s^*$        &  $01^-$           &     $2140\pm4$     &   2130    &   2112   &   2112             \\
$\eta_c$       &  $00^-$           &     $2912\pm5$     &   2970    &   2990   &   2980             \\
$J/\Psi$       &  $01^-$           &     $3102\pm4$     &   3100    &   3097   &   3097             \\
$B^0$          &  $\frac{1}{2}0^-$ &     $5259\pm5$     &   5310    &   5281   &   5280             \\
$B^*$          &  $\frac{1}{2}1^-$ &     $5301\pm4$     &   5370    &   5321   &   5325             \\
$B_s^0$        &  $00^-$           &     $5377\pm5$     &   5390    &   5355   &   5366             \\
$B_s^*$        &  $01^-$           &     $5430\pm4$     &   5450    &   5400   &   5416             \\
$B_c$          &  $00^-$           &     $6261\pm7$     &   6270    &   6277   &   6277             \\
$B_c^*$        &  $01^-$           &     $6357\pm4$     &   6340    &   ...    &   ...              \\
$\eta_b$       &  $00^-$           &     $9441\pm8$     &   9400    &   9454   &   9391             \\
$\Upsilon(1S)$ &  $01^-$           &     $9546\pm5$     &   9460    &   9505   &   9460             \\
\toprule[0.8pt]
\end{tabular}
\end{table}
Meson spectrum have been also studied in other different quark models~\cite{chiralmeson,rel-meson}. The spectrum from the light-pseudoscalar and vector mesons to bottomonium are also investigated in a nonrelativistic quark model (\textbf{17} free parameters) with one gluon exchange potential, a screened confinement and one boson exchange~\cite{chiralmeson}. The mesons from the $\pi$ to $\Upsilon$ can be described in a relativized quark model (\textbf{14} free parameters) with a universal one gluon exchange plus a linear confining potential motivated by QCD~\cite{rel-meson}. For comparison, the results of other two models are also listed in the Table \ref{mesons}. Objectively speaking, the other two models can describe the meson spectra a little better than our model, the main reason of which is that the number of the free parameters in our model is much less than those in the two models because the ACFTM has just $\textbf{8}$ adjustable parameters. On the whole, the ACFTM can describe meson spectrum the point of view of the model. In general, it is hard to exactly produce a large amount of states in the quark model calculation with limited number of parameters. The more parameters the model has, the more accurate it is. One does not expect to introduce too many free parameters to improve the accuracy of meson spectrum at the expense of reducing the prediction ability of the model.

In addition, it can be found that the non-relativistic quark model and relativistic one are equivalent, which can give a reasonable meson spectrum. A great deal of early researches on meson spectra have been devoted to compare the equivalence of various types of quark models~\cite{three-models}. Phenomenological model researches on multiquark states and hadron-hadron interactions hope that the good equivalence found between relativistic and nonrelativistic meson spectra persists for multi-quark systems. In fact, norelativistic quark models have been successfully applied to baryon-baryon interactions and new hadrons observed in experiments up to now~\cite{NNNY,qdcsm, newhadrons}.

Although it is generally recognized that the models with relativistic dynamics are more rigorous from the theoretical point of view, all relativistic quark models had to face the technical difficulty, an endemic problem, of separating the centre of mass motion. In contrast, nonrelativistic quark models can cope with the centre of mass motion and also be more easily extended to multibody dynamics than relativistic ones.

\section{wavefunctions of the doubly heavy states}

The properties of the doubly heavy tetraquark states can be obtained using a complete wavefunction which includes all possible flavor-spin-color-spatial channels that contribute to a given well defined parity, isospin, and total angular momentum. Within the framework of the diquark-antidiquark configuration, the trial wavefunction of the doubly heavy tetraquark state $[QQ][\bar{q}\bar{q}]$ can be constructed as a sum of the following direct products of color $\chi_c$, isospin $\eta_i$, spin $\chi_s$ and spatial $\phi$ terms
\begin{eqnarray}
\Phi^{[QQ][\bar{q}\bar{q}]}_{IM_IJM_J} &=& \sum_{\alpha}\xi_{\alpha}\left[\left[\left[\phi_{l_am_a}^G(\mathbf{r})\chi_{s_a}\right]^{[QQ]}_{J_aM_{J_a}}
\left[\phi_{l_bm_b}^G(\mathbf{R})\right.\right.\right.\nonumber\\
& \times & \left.\left.\left.\chi_{s_b}\right]^{[\bar{q}\bar{q}]}_{J_bM_{J_b}}\right ]_{J_{ab}M_{J_{ab}}}^{[QQ][\bar{q}\bar{q}]}
\phi^G_{l_cm_c}(\mathbf{X})\right]^{[QQ][\bar{q}\bar{q}]}_{JM_J}\\
& \times &
\left[\eta_{i_a}^{[QQ]}\eta_{i_b}^{[\bar{q}\bar{q}]}\right]_{IM_I}^{[QQ][\bar{q}\bar{q}]}
\left[\chi_{c_a}^{[QQ]}\chi_{c_b}^{[\bar{q}\bar{q}]}\right]_{CW_C}^{[QQ][\bar{q}\bar{q}]},
\nonumber
\end{eqnarray}
The subscripts $a$ and $b$ represent the diquark $[QQ]$ and antidiquark $[\bar{q}\bar{q}]$, respectively. The summering index $\alpha$ stands for all possible flavor-spin-color-spatial intermediate quantum numbers.

The relative spatial coordinates $\mathbf{r}$, $\mathbf{R}$ and $\mathbf{X}$ and center of mass $\mathbf{R}_c$ in the tetraquark state $[QQ][\bar{q}\bar{q}]$ can be defined as,
\begin{eqnarray}
\mathbf{r}&=&\mathbf{r}_1-\mathbf{r}_2,~~~\mathbf{R}=\mathbf{r}_3-\mathbf{r}_4 \nonumber\\
\mathbf{X}&=&\frac{m_1\mathbf{r}_1+m_2\mathbf{r}_2}{m_1+m_2}-\frac{m_3\mathbf{r}_3+m_4\mathbf{r}_4}{m_3+m_4},\\
\mathbf{R}_c&=&\frac{m_1\mathbf{r}_1+m_2\mathbf{r}_2+m_3\mathbf{r}_3+m_4\mathbf{r}_4}{m_1+m_2+m_3+m_4}.\nonumber
\end{eqnarray}
The corresponding angular excitations of three relative motions are, respectively, $l_a$, $l_b$ and $l_c$. The parity of the doubly heavy tetraquark states $[QQ][\bar{q}\bar{q}]$ can therefore be expressed in terms of the relative orbital angular momenta associated with the Jacobi coordinates as $P=(-1)^{l_a+l_b+l_c}$. It is worth mentioning that this set of coordinate is only a possible choice of many coordinates and however most propitious to describe the correlation of two quarks (antiquark) in the diquark (antidiqurk) and construct the symmetry of identical particles. In order to obtain a reliable solution of few-body problem, a high precision numerical method is indispensable. The Gaussian Expansion Method (GEM)~\cite{GEM}, which has been proven to be rather powerful to solve few-body problem, is therefore used to study four-quark systems in the present work. According to the GEM, any relative motion wave function can be written as,
\begin{eqnarray}
\phi^G_{lm}(\mathbf{z})=\sum_{n=1}^{n_{max}}c_{n}N_{nl}z^{l}e^{-\nu_{n}z^2}Y_{lm}(\hat{\mathbf{z}})
\end{eqnarray}
More details of the relative motion wave functions can be found in the paper~\cite{GEM}.

The color representation of the diquark maybe antisymmetrical $[QQ]_{\bar{\mathbf{3}}_c}$ or symmetrical $[QQ]_{\mathbf{6}_c}$, whereas that of the antidiquark maybe antisymmetrical $[\bar{q}\bar{q}]_{\mathbf{3}_c}$ or symmetrical $[\bar{q}\bar{q}]_{\bar{\mathbf{6}}_c}$. Coupling the diquark and the antidiquark into an overall color singlet according to color coupling rule have two ways:  $\left[[QQ]_{\bar{\mathbf{3}}_c}[\bar{q}\bar{q}]_{\mathbf{3}_c}\right]_{\mathbf{1}}$ (good diquark) and $\left[[QQ]_{\mathbf{6}_c}[\bar{q}\bar{q}]_{\bar{\mathbf{6}}_c}\right]_{\mathbf{1}}$ (bad diquark). In general, the interaction in the good diquark is attractive, whereas the interaction in the bad diquark is repulsive. Anyway, a real physical state should be their mixture because of the coupling between two color configurations.

The spin of the diquark $[QQ]$ is coupled to $s_a$ and that of the antiquarks $[\bar{q}\bar{q}]$ to $s_b$. The total spin wave function of the doubly heavy tetraquark state $[QQ][\bar{q}\bar{q}]$ can be written as $S=s_a\oplus s_b$. Then we have the following basis vectors as a function of the total
spin $S$.
\begin{eqnarray}
S=\left\{
\begin{array}{ll}
\mbox{$0=1\oplus1$ and $0\oplus0$}\\
\mbox{$1=1\oplus1$, $1\oplus0$, and $0\oplus1$}\\
\mbox{$2=1\oplus1$}\\
\end{array}
\right. ,
\end{eqnarray}

With respect to the flavor wavefunction, we only consider $SU_f(2)$ symmetry in the present work. The quarks, $s$, $c$ and $b$, have isospin zero so they do not contribute to the total isospin. The flavor wave functions of the antidiquark consisting of $\bar{u}$ and $\bar{d}$ quarks are similar to those of spin.

Taking all degrees of freedom of identical particles in the diquark (antidiquark) into account, the Pauli principle must be satisfied by imposing some restrictions on the quantum numbers of the diquark (antidiquark). Such as the color-antisymmetrical tetraquark state $[cc]_{\bar{\mathbf{3}}_c}[\bar{u}\bar{d}]_{\mathbf{3}_c}$, the quantum numbers must satisfy the relations $(-1)^{l_a+i_a+s_a}=-1$ and $(-1)^{l_b+i_b+s_b}=1$. But for the color-symmetrical tetraquark state $[cc]_{\mathbf{6}_c}[\bar{u}\bar{d}]_{\bar{\mathbf{6}}_c}$, the quantum numbers must satisfy the relations $(-1)^{l_a+i_a+s_a}=1$ and $(-1)^{l_b+i_b+s_b}=-1$. On the contrary, the situation of non-identical particles is extremely simple because of no any restrictions.

\section{numerical results and analysis}

The converged numerical results of the doubly heavy tetraquark states $[QQ][\bar{q}\bar{q}]$ within the framework of the ACFTM can be obtained through solving the four-body Schr\"{o}dinger equation with the Rayleigh-Ritz variational principle,
\begin{eqnarray}
(H_4-E_4)\Phi^{[QQ][\bar{q}\bar{q}]}_{IM_IJM_J}=0.
\end{eqnarray}
A tetraquark state should be stable against strong interaction if its energy lies below all possible two-meson thresholds. We express the lowest threshold of the doubly heavy tetraquark $[QQ][\bar{q}\bar{q}]$ as $T^{min}_{M_1M_2}$, where $M_1$ and $M_2$ stand for two $Q\bar{q}$ mesons. The binding energy of the doubly heavy tetraquark states can be therefore defined as
\begin{eqnarray}
E_b=E_4-T^{min}_{M_1M_2}
\end{eqnarray}
to identify whether or not the tetraquark state is stable against strong interactions. The procedure can greatly reduce the influence of inaccurate meson spectra coming from the parameters on the binding energies by the theoretical difference between the energy of the tetraquark states and that of two mesons. If $E_b\geq0$, the tetraquark state can fall apart into two mesons via strong interactions. If $E_b<0$, the strong decay into two mesons is forbidden and therefore the decay must be weak or electromagnetic interaction.
\begin{table}[ht]
\caption{The energies of the doubly heavy tetraquark states $[QQ][\bar{q}\bar{q}]$, masses unit in MeV.} \label{spectrum}
\begin{tabular}{ccccccccccc}
\toprule[0.8pt]
~~Flavor~~                 &       ~$IJ^{P}$~~    & $n^{2S+1}L_J$ &  ~~~~Masses~~~~ &  ~~$T^{min}_{M_1M_2}$~~  & ~~$E_b$~~ \\
\toprule[0.8pt]
                           &        $01^{+}$      &    $0^3S_1$   &   $3719\pm12$   &         ${DD^*}$         &  $-150$   \\
                           &        $01^{-}$      &    $0^1P_1$   &   $3931\pm12$   &         ${DD}$           &  $197$    \\
$[cc][\bar{u}\bar{d}]$     &        $10^{+}$      &    $0^1S_0$   &   $3962\pm8$    &         ${DD}$           &  228      \\
                           &        $11^{+}$      &    $0^3S_1$   &   $4017\pm7$    &         ${DD^*}$         &  148      \\
                           &        $12^{+}$      &    $0^5S_2$   &   $4013\pm7$    &         ${D^*D^*}$       &  9        \\
\toprule[0.8pt]
                           &        $00^{+}$      &    $0^1S_0$   &   $6990\pm12$   &         ${DB}$           &  $-136$   \\
                           &        $01^{+}$      &    $0^3S_1$   &   $6997\pm12$   &         ${DB^*}$         &  $-171$   \\
                           &        $02^{+}$      &    $0^5S_2$   &   $7321\pm7$    &         ${D^*B^*}$       &  18       \\
$[bc][\bar{u}\bar{d}]$     &        $01^{-}$      &    $0^1P_1$   &   $7154\pm9$    &         ${DB}$           &  28       \\
                           &        $10^{+}$      &    $0^1S_0$   &   $7270\pm8$    &         ${DB}$           &  144      \\
                           &        $11^{+}$      &    $0^3S_1$   &   $7283\pm8$    &         ${DB^*}$         &  115      \\
                           &        $12^{+}$      &    $0^5S_2$   &   $7299\pm7$    &         ${D^*B^*}$       &  $-4$     \\
\toprule[0.8pt]
                           &        $01^{+}$      &    $0^3S_1$   &   $10282\pm12$  &         ${BB^*}$         &  $-278$   \\
                           &        $01^{-}$      &    $0^1P_1$   &   $10404\pm11$  &         ${BB}$           &  $-114$   \\
$[bb][\bar{u}\bar{d}]$     &        $10^{+}$      &    $0^1S_0$   &   $10558\pm7$   &         ${BB}$           &  40       \\
                           &        $11^{+}$      &    $0^3S_3$   &   $10586\pm7$   &         ${BB^*}$         &  26       \\
                           &        $12^{+}$      &    $0^5S_2$   &   $10572\pm7$   &         ${B^*B^*}$       &  $-30$    \\
\toprule[0.8pt]
                           &  $\frac{1}{2}0^{+}$  &    $0^1S_0$   &   $4121\pm8$    &         ${DD_s}$         &  282      \\
$[cc][\bar{u}\bar{s}]$     &  $\frac{1}{2}1^{+}$  &    $0^3S_1$   &   $4068\pm9$    &         ${D^*D_s}$       &  94       \\
                           &  $\frac{1}{2}2^{+}$  &    $0^5S_2$   &   $4177\pm7$    &         ${D^*D_s^*}$     &  35       \\
\toprule[0.8pt]
                           &  $\frac{1}{2}0^{+}$  &    $0^1S_0$   &   $7339\pm9$    &         ${D_sB}$         &  108      \\
$[bc][\bar{u}\bar{s}]$     &  $\frac{1}{2}1^{+}$  &    $0^3S_1$   &   $7356\pm9$    &         ${D_sB^*}$       &  $83$     \\
                           &  $\frac{1}{2}2^{+}$  &    $0^5S_2$   &   $7455\pm7$    &         ${D^*B_s^*}$     &  23       \\
\toprule[0.8pt]
                           &  $\frac{1}{2}0^{+}$  &    $0^1S_0$   &   $10716\pm7$   &         ${BB_s}$         &  80       \\
$[bb][\bar{u}\bar{s}]$     &  $\frac{1}{2}1^{+}$  &    $0^3S_1$   &   $10629\pm9$   &         ${B^*B_s}$       &  $-49$    \\
                           &  $\frac{1}{2}2^{+}$  &    $0^5S_2$   &   $10734\pm7$   &         ${B^*B_s^*}$     &  3        \\
\toprule[0.8pt]
                           &        $00^{+}$      &    $0^1S_0$   &   $4279\pm8$    &         ${D_sD_s}$       &  335      \\
$[cc][\bar{s}\bar{s}]$     &        $01^{+}$      &    $0^3S_1$   &   $4312\pm7$    &         ${D_sD_s^*}$     &  193      \\
                           &        $02^{+}$      &    $0^5S_2$   &   $4328\pm7$    &         ${D^*_sD_s^*}$   &  48       \\
\toprule[0.8pt]
                           &        $00^{+}$      &    $0^1S_0$   &   $7582\pm7$    &         ${D_sB_s}$       &  232      \\
$[bc][\bar{s}\bar{s}]$     &        $01^{+}$      &    $0^3S_1$   &   $7590\pm7$    &         ${D_sB_s^*}$     &  188      \\
                           &        $02^{+}$      &    $0^5S_2$   &   $7611\pm7$    &         ${D^*_sB_s^*}$   &  41       \\
\toprule[0.8pt]
                           &        $00^{+}$      &    $0^1S_0$   &   $10866\pm7$   &         ${B_sB_s}$       &  112      \\
$[bb][\bar{s}\bar{s}]$     &        $01^{+}$      &    $0^3S_1$   &   $10875\pm7$   &         ${B_sB_s^*}$     &  68       \\
                           &        $02^{+}$      &    $0^5S_2$   &   $10882\pm7$   &         ${B^*_sB_s^*}$   &  22       \\
\toprule[0.8pt]
\end{tabular}
\end{table}

In the following, we discuss the properties of the doubly heavy tetraquark states $[QQ][\bar{q}\bar{q}]$ to search for all possible stable states against strong interactions in the ACFTM. In order to obtain the lowest states with positive parity, we assume that three relative motions are in a relative S-wave in the doubly heavy states. In the case of the lowest states with negative parity, we assume that the angular excitation of the relative motion occur in not $l_b$ and $l_c$ but $l_a$, namely $l_a=1$, $l_b=l_c=0$. The reason is that the angular excitation in the diquark $[QQ]$ bring a kinetic energy as possible as small into the excited states, which contributes to the stability of the doubly heavy tetraquark states $[QQ][\bar{q}\bar{q}]$. The ACFTM predictions on the lowest energies of the doubly heavy tetraquark states $[QQ][\bar{q}\bar{q}]$ with a set of given $IJ^P$ are presented in Table \ref{spectrum}. A first glance gives a conclusion that there exists seven bound states with positive parity, the states $[cc][\bar{u}\bar{d}]$, $[bc][\bar{u}\bar{d}]$ and $[bb][\bar{u}\bar{d}]$ with $01^+$, the states $[bc][\bar{u}\bar{d}]]$ and $[bb][\bar{u}\bar{d}]$ with $12^+$, the strange state $[bb][\bar{u}\bar{s}]$ with $\frac{1}{2}1^+$ and the state $[bc][ud]$ with $00^+$, and one negative parity state $[bb][\bar{u}\bar{d}]$ with $01^-$. Other doubly heavy tetraquark states lie above the corresponding lowest threshold within the framework of the ACFTM and should therefore  decay very rapidly through the fall-apart mechanism of the color flux-tubes.
\begin{table*}
\caption{The stable doubly heavy tetraquark states $[QQ][\bar{q}\bar{q}]$ in various methods, two results in Ref.~\cite{pepin} for two different sets of parameters $C_1$ and $C_2$, unit in MeV.}\label{comparison}
\begin{tabular}{cccccccccccccccccccc}
\toprule[0.8pt]
  States&&&~~Ours~~&&&&&&&Others&&\\
Flavor&~$IJ^P$~&$T^{min}_{M_1M_2}$&ACFTM&\cite{karliner} &~~~\cite{Eichten}~~~&~\cite{Francis}~&~~~~\cite{bicudo}~~~~&\cite{cqmpark} &~~~\cite{ebert}~~~&\cite{ccqqvij}&\cite{pepin}&\cite{sakai}&\cite{valcarce}&~~~\cite{park}~~~&\cite{semay} \\
\toprule[0.8pt]
$[cc][\bar{u}\bar{d}]$&$01^+$&$DD^*$            & $-150$ &  7     &  ...  &  ...   &  ...              &  ...   &  64     &$-129$ & $-185,-332$  &    ...     & $-76$ & $100$  & 19      \\
$[bc][\bar{u}\bar{d}]$&$00^+$&$DB$              & $-136$ & $-11$  &  ...  &  ...   &  ...              &  ...   &  95     & ...   & ...          & $-[20,60]$ & ...   & ...    & 11      \\
$[bc][\bar{u}\bar{d}]$&$01^+$&$DB^*$            & $-171$ &  ...   &  ...  &  ...   &  ...              &  ...   &  56     & ...   & ...          & $-[20,60]$ & ...   & ...    & 1       \\
$[bc][\bar{u}\bar{d}]$&$12^+$&$D^*B^*$          & $-4$   &  ...   &  ...  &  ...   &  ...              &  ...   &  90     & ...   & ...          & $-[20,60]$ & ...   & ...    & ...     \\
$[bb][\bar{u}\bar{d}]$&$01^+$&$BB^*$            & $-278$ & $-215$ & $-121$& $-189$ & $-59^{+30}_{-38}$ & $-121$ &  $-102$ &$-341$ & $-226,-497$  & ...        & $-214$& $-100$ & $-131$  \\
$[bb][\bar{u}\bar{d}]$&$12^+$&$B^*B^*$          & $-30$  &  ...   &  ...  &  ...   &  ...              &  ...   &  23     &65     & ...          & ...        &  1    & ...    & 30      \\
$[bb][\bar{u}\bar{s}]$&$\frac{1}{2}1^+$&$B^*B_s$& $-49$  &  ...   & $-48$ &  $-98  $ &...              &  $-7$  &  13     &...    & ...          & ...        & ...   & ...    & $-40$   \\
$[bb][\bar{u}\bar{d}]$&$01^-$&$BB$              & $-114$ &  ...   &  ...  &  ...   &  ...              &  ...   &  ...    &...    & ...          & ...        & 11    & ...    & ...     \\
\toprule[0.8pt]
\end{tabular}
\end{table*}
\begin{table}
\caption{The energies of all stable states with the color configurations $\left[[QQ]_{\bar{\mathbf{3}}_c}[\bar{q}\bar{q}]_{\mathbf{3}_c}\right]_{\mathbf{1}}$ and  $\left[[QQ]_{\mathbf{6}_c}[\bar{q}\bar{q}]_{\bar{\mathbf{6}}_c}\right]_{\mathbf{1}}$ in the ACFTM, unit in MeV.}\label{color configurations}
\begin{tabular}{cccccccccc}
\toprule[0.8pt]
~~~Flavor~~~&  ~$IJ^P$~  &  $~~~~~\bar{\mathbf{3}}_c\otimes\mathbf{3}_c$~~~~~&~~~$\mathbf{6}_c\otimes\bar{\mathbf{6}}_c$~~~~&~~Coupling~~~  \\
$[cc][\bar{u}\bar{d}]$  &    $01^+$          &   $3731\pm12$  &   $4007\pm8$   &   $3719\pm12$   \\
$[bc][\bar{u}\bar{d}]$  &    $00^+$          &   $6996\pm12$  &   $7262\pm8$   &   $6990\pm12$   \\
$[bc][\bar{u}\bar{d}]$  &    $01^+$          &   $7003\pm12$  &   $7304\pm7$   &   $6997\pm12$   \\
$[bc][\bar{u}\bar{d}]$  &    $12^+$          &   $7299\pm7$   &      ...       &   $7299\pm7$    \\
$[bb][\bar{u}\bar{d}]$  &    $01^+$          &   $10283\pm12$ &   $10583\pm8$  &   $10282\pm12$  \\
$[bb][\bar{u}\bar{d}]$  &    $12^+$          &   $10572\pm7$  &      ...       &   $10572\pm7$   \\
$[bb][\bar{q}'\bar{s}]$ &  $\frac{1}{2}1^+$  &   $10629\pm9$  &   $10721\pm8$  &   $10629\pm9$   \\
$[bb][\bar{u}\bar{d}]$  &    $01^-$          &   $10404\pm12$ &   $10847\pm7$  &   $10404\pm12$  \\
\toprule[0.8pt]
\end{tabular}
\caption{The average distance $\langle\mathbf{r}_{ij}^2\rangle^{\frac{1}{2}}$ between the $i$-th and $j$-th particle in the stable states, unit in fm.}\label{rms}
\begin{tabular}{cccccccccc}
\toprule[0.8pt]
~~Flavor~~&$IJ^P$&$E_b$  &$\langle\mathbf{r}_{12}^2\rangle^{\frac{1}{2}}$$\langle\mathbf{r}_{34}^2\rangle^{\frac{1}{2}}$$\langle\mathbf{r}_{24}^2\rangle^{\frac{1}{2}}$&
          $\langle\mathbf{r}_{13}^2\rangle^{\frac{1}{2}}$$\langle\mathbf{r}_{14}^2\rangle^{\frac{1}{2}}$$\langle\mathbf{r}_{23}^2\rangle^{\frac{1}{2}}$~\\
$[cc][\bar{u}\bar{d}]$&$01^+$             &  $-150$  &  0.65~~  0.78~~  0.91  &  0.91~~  0.91~~  0.91  \\
$[bc][\bar{u}\bar{d}]$&$00^+$             &  $-136$  &  0.53~~  0.78~~  0.91  &  0.83~~  0.83~~  0.91  \\
$[bc][\bar{u}\bar{d}]$&$01^+$             &  $-171$  &  0.55~~  0.78~~  0.92  &  0.84~~  0.84~~  0.92  \\
$[bc][\bar{u}\bar{d}]$&$12^+$             &  $-4$    &  0.56~~  1.13~~  1.06  &  0.98~~  0.98~~  1.06  \\
$[bb][\bar{u}\bar{d}]$&$01^+$             &  $-278$  &  0.42~~  0.77~~  0.84  &  0.84~~  0.84~~  0.84  \\
$[bb][\bar{u}\bar{d}]$&$12^+$             &  $-30$   &  0.42~~  1.13~~  0.98  &  0.98~~  0.98~~  0.98  \\
$[bb][\bar{u}\bar{s}]$&$\frac{1}{2}1^+$   &  $-49$   &  0.42~~  0.89~~  0.90  &  0.76~~  0.76~~  0.90  \\
$[bb][\bar{u}\bar{d}]$&$01^-$             &  $-114$  &  0.65~~  0.77~~  0.89  &  0.89~~  0.89~~  0.89  \\
\toprule[0.8pt]
\end{tabular}
\end{table}
\begin{table*}
\caption{The contributions of the various parts of the Hamiltonian in the ACFTM to the masses and the binding energies $E_B$ of the stable states
$[QQ][\bar{q}\bar{q}]$, where $V^{B}$, $V^{\sigma}$, $V^C$, $V^{cm}$, $E_k$, and $V^{clb}$ represent one Goldstone boson exchange, $\sigma$-meson exchange, confinement, color-magnetic interaction, kinetic energy and Coulomb items, respectively, unit in MeV.}\label{contributions}
\begin{tabular}{ccccccccc||cccccccccccc}
\toprule[0.8pt]
~Flavor~&~$IJ^P$~&Masses&~~~$V^{\sigma}$~~~&~~$V^{B}$~~&~~~$V^C$~~~&~~$V^{cm}$~~&~~~$E_k$~~~&~~$V^{clb}$~~&&~~$E_b$~~&$~~\Delta V^{\sigma}~~$&~$\Delta V^{B}$~&$~~\Delta V^C$~~&$~\Delta V^{cm}$~&~$\Delta E_k$~&$\Delta V^{clb}$ \\
$[cc][\bar{u}\bar{d}]$ & $01^+$           &$3719\pm12$ &$-35$&$-223$&173&$-236$&953&$-676$ && $-150$&$-35$&$-223$&$-73$ &$-98$ & 174 &105\\
$[bc][\bar{u}\bar{d}]$ & $00^+$           &$6990\pm12$ &$-37$&$-245$&153&$-261$&993&$-711$ && $-136$&$-37$&$-245$&$-61$ &$-65$ & 147 &125\\
$[bc][\bar{u}\bar{d}]$ & $01^+$           &$6997\pm12$ &$-37$&$-246$&155&$-253$&983&$-703$ && $-171$&$-37$&$-246$&$-75$ &$-103$& 196 &93 \\
$[bc][\bar{u}\bar{d}]$ & $12^+$           &$7299\pm7$  &$-15$&$-10$ &242& 30   &506&$-552$ && $-4$  &$-15$&$-10$ &$-34$ &  0   &$-68$&123\\
$[bb][\bar{u}\bar{d}]$ & $01^+$           &$10282\pm12$&$-36$&$-228$&140&$-236$&928&$-719$ && $-278$&$-36$&$-228$&$-104$&$-208$& 288 &11 \\
$[bb][\bar{u}\bar{d}]$ & $12^+$           &$10572\pm7$ &$-15$&$-10$ &222& 27   &490&$-574$ && $-30$ &$-15$&$-10$ &$-38$ & 9    &$-92$&116\\
$[bb][\bar{u}\bar{s}]$ & $\frac{1}{2}1^+$ &$10629\pm11$&$-27$&$-10$ &151&$-98$ &605&$-654$ && $-49$ &$-27$&$-10$ &$-49$ &$-54$ &$-41$&132\\
$[bb][\bar{u}\bar{d}]$ & $01^-$           &$10404\pm9$ &$-37$&$-244$&170&$-253$&962&$-626$ && $-114$&$-37$&$-244$&$-59$ &$-179$& 262 &144\\
\toprule[0.8pt]
\end{tabular}
\end{table*}
\begin{table}
\caption{The energies of the stable states in the two models with fourbody and twobody confinement potential and the difference between two central values, unit in MeV.}\label{diff}
\begin{tabular}{cccccccccc}
\toprule[0.8pt]
~~~Flavor~~~            &   $IJ^P$          &  ~~Two-body~~  & ~~~Four-body~~~&  ~Difference~  \\
$[cc][\bar{u}\bar{d}]$  &   $01^+$          &   $3817\pm11$  &  $3719\pm12$   &   $98$        \\
$[bc][\bar{u}\bar{d}]$  &   $00^+$          &   $7096\pm11$  &  $6990\pm12$   &   $106$       \\
$[bc][\bar{u}\bar{d}]$  &   $01^+$          &   $7109\pm11$  &  $6997\pm12$   &   $102$       \\
$[bc][\bar{u}\bar{d}]$  &   $12^+$          &   $7440\pm7$   &  $7299\pm7$    &   $141$       \\
$[bb][\bar{u}\bar{d}]$  &   $01^+$          &   $10355\pm11$ &  $10282\pm12$  &   $73$        \\
$[bb][\bar{u}\bar{d}]$  &   $12^+$          &   $10698\pm7$  &  $10572\pm7$   &   $126$       \\
$[bb][\bar{q}'\bar{s}]$ &  $\frac{1}{2}1^+$ &   $10729\pm8$  &  $10629\pm9$   &   $100$       \\
$[bb][\bar{u}\bar{d}]$  &   $01^-$          &   $10508\pm11$ &  $10404\pm12$  &   $104$       \\
\toprule[0.8pt]
\end{tabular}
\end{table}

The binding energies of the doubly heavy tetraquark states within various theoretical methods are presented in Table \ref{comparison}, in which ``...'' represents
that the corresponding state was not researched by authors. It is extremely obvious that the state $[bb][\bar{u}\bar{d}]$ with $01^+$ has a distinguished strong binding, above 100 MeV, in the absolutely majority of work and must therefore be the most promising stable doubly heavy tetraquark state against dissociation into two heavy-light mesons via strong interaction. Its strange partner, $[bb][\bar{u}\bar{s}]$ with $\frac{1}{2}1^+$, has also a binding energies from a few to dozens of MeV in all of investigations with exception of Ebert's work~\cite{ebert}, which lies slightly, about 13 MeV, above the $B^*B_s$ threshold. In this way, the state $[bb][\bar{u}\bar{s}]$ with $\frac{1}{2}1^+$ stands a good chance of existence as a bound state. It is strongly suggested that the two extremely possible stable states against strong interactions should be explored in experiments in the near future.

In addition to the two doubly heavy tetraquark states $[bb][\bar{u}\bar{d}]$ with $01^+$ and $[bb][\bar{u}\bar{s}]$ with $\frac{1}{2}1^+$, the existent of other states in Table \ref{comparison} as stable states against strong interactions are obviously model dependent. The state $[cc][\bar{u}{d}]$ with $01^+$ lies below, greater than 100 MeV, the threshold $DD^*$  only in the ACFTM and the chiral quark models~\cite{ccqqvij,pepin}. Other results on the state are higher than the threshold $DD^*$. In the case of the states $[bc][\bar{u}\bar{d}]$ with $00^+$, $01^+$ and $12^+$, Sakai et al described them as $D^{(*)}B^{(*)}$ molecule states with binding energies about 20--60 MeV~\cite{sakai}. QCD sum rule research indicated that the extracted masses for both the scalar and axial vector $[bc][\bar{q}\bar{q}]$ tetraquark states are also below the open-flavor thresholds $DB$ and $DB^*$~\cite{qcdsumbcqq}. Lattice QCD study shown the existence of a strong-interaction-stable tetraquark $[bb][\bar{u}\bar{d}]$ with $01^+$  below $DB^*$ threshold in the range of 15 to 61 MeV~\cite{francis}. In the ACFTM, the states $[bc][\bar{u}\bar{d}]$ with $00^+$ and $01^+$ can be depicted as deeply bound states with binding energies $136$ MeV and $171$ MeV, respectively. The state $[bc][\bar{u}\bar{d}]$ with $12^+$ as a bound state have a slight binding, about 4 MeV. In this way, our conclusion on the three heavy states $[bc][\bar{u}\bar{d}]$ is qualitatively consistent with that of Sakai. Karliner also predicted that the state $[bc][\bar{u}\bar{d}]$ with $00^+$ lies below the threshold $DB$ about 11 MeV~\cite{karliner}. The heavy state $[bb][\bar{u}\bar{d}]$, the partner of $[bc][\bar{u}\bar{d}]$ with $12^+$, can exist as a stable state with binding energy about 30 MeV, which is not supported by existing results on the doubly heavy tetraquark states so far. With respect to the state $[bb][\bar{u}\bar{d}]$ with $1^-$, the ACFTM predicts that it lies below the $BB$ threshold about 114 MeV. The energy of this state is higher, just 1 MeV, than the threshold in Ref.~\cite{valcarce}. Very recently, Pflaumer et al predicted the doubly heavy tetraquark state $[bb][\bar{u}\bar{d}]$ with $1^-$ as a resonance higher 17 MeV than the $BB$ threshold applying lattice QCD potentials~\cite{pflaumer}. In general, the heavy states $[QQ][\bar{u}\bar{d}]$ with $I=0$ are easier than the states with $I=1$ to form bound states in the ACFTM.

All possible stable doubly heavy tetraquark states should be, in general, the admixture of the two color configurations $\left[[QQ]_{\bar{\mathbf{3}}_c}[\bar{q}\bar{q}]_{\mathbf{3}_c}\right]_{\mathbf{1}}$ and  $\left[[QQ]_{\mathbf{6}_c}[\bar{q}\bar{q}]_{\bar{\mathbf{6}}_c}\right]_{\mathbf{1}}$ under the diquark-antidiquark picture as a working hypothesis. Theoretically, the magnitude of their mixing through color-magnetic interaction is governed by the order of $\frac{1}{m_im_j}$. Special attention is therefore payed to the role quantitatively played by the two color configurations in the ACFTM. The energies of all possible stable doubly heavy tetraquark states with the two color configurations and their coupling results are given in Table \ref{color configurations}. It can be found that the configuration $\left[[QQ]_{\bar{\mathbf{3}}_c}[\bar{q}\bar{q}]_{\mathbf{3}_c}\right]_{\mathbf{1}}$ dominates the energy of the doubly heavy tetraquark states. The
mixing effect pushes the energy of the states down a little comparing with that of the configuration $\left[[QQ]_{\bar{\mathbf{3}}_c}[\bar{q}\bar{q}]_{\mathbf{3}_c}\right]_{\mathbf{1}}$. The heavier the heavy quark pair $[QQ]$, the stronger the effect. For the $[cc]$ and $[bc]$ sections, it is just more than ten MeV and several MeV, respectively. In the $[bb]$ section, the color configuration $\left[[bb]_{\mathbf{6}_c}[\bar{q}\bar{q}]_{\bar{\mathbf{6}}_c}\right]_{\mathbf{1}}$ has almost no any effect on the masses of the states and can be ignored in the ACFTM. Therefore, the color configuration $\left[[QQ]_{\bar{\mathbf{3}}_c}[\bar{q}\bar{q}]_{\mathbf{3}_c}\right]_{\mathbf{1}}$ absolutely dominates the behavior of the doubly heavy tetraquark states in the course of investigation on their properties~\cite{ccqqvij,park}. However, the color configuration $\left[qq]_{\mathbf{6}_c}[\bar{q}\bar{q}]_{\bar{\mathbf{6}}_c}\right]_{\mathbf{1}}$ must be taken into accounted in the researches on the light tetraquark states~\cite{ccqqvij}.

Regarding to the stable doubly heavy tetraquark states $\left[[bb]_{\bar{\mathbf{3}}_c}[\bar{u}\bar{d}]_{\mathbf{3}_c}\right]_{\mathbf{1}}$ and $\left[[cc]_{\bar{\mathbf{3}}_c}[\bar{u}\bar{d}]_{\mathbf{3}_c}\right]_{\mathbf{1}}$ with $01^+$, the two heavy diquarks $[bb]_{\bar{\mathbf{3}}}$ and $[cc]_{\bar{\mathbf{3}}}$ must have spin one because the flavor and orbit are symmetrical, the color-spin-orbit-isospin combination is $(c_a,s_a,l_a,i_a)=(\bar{\mathbf{3}}_c,1,0,0)$.
The antidiquark $[\bar{u}\bar{d}]_{\mathbf{3}}$ couples into spin and isospin zero, the color-spin-orbit-isospin combination is $(c_b,s_b,l_b,i_b)=(\mathbf{3}_c,0,0,0)$. For the heavy diquarks $[bb]_{\bar{\mathbf{3}}}$ and $[cc]_{\bar{\mathbf{3}}}$, the color-magnetic interaction is therefore weak repulsive. However, the large masses admit two heavy quarks to approach each other as short as possible because kinetic energy is inversely proportional to the quark mass. Meanwhile, the Coulomb interaction is attractive in the diquark $[QQ]_{\bar{\mathbf{3}}}$. The heavier the heavy quark, the stronger the Coulomb interaction, the shorter the distance, see Table \ref{rms}. In the limit of heavy quark, the diquark $[QQ]_{\bar{\mathbf{3}}}$ gradually shrink into a pointlike particle, which is qualitatively consistently with the conclusion in Quigg's work~\cite{heavy}. With the exception of attractive Coulomb interaction, there exists strong attractive interactions in the antidiquark $[\bar{u}\bar{d}]_{\mathbf{3}_c}$ with spin and isospin zero generated through one-Goldstone-boson-exchange (mainly $\pi$) and color-magnetic interaction. This conclusion is also hold for the state $\left[[bc]_{\bar{\mathbf{3}}_c}[\bar{u}\bar{d}]_{\mathbf{3}_c}\right]_{\mathbf{1}}$ with $01^+$. In this way, the interaction in the doubly heavy states $[QQ][\bar{u}\bar{d}]$ with $01^+$ become strong gradually with the increase of the mass ratio $\frac{m_Q}{m_{\bar{q}}}$, which was pointed out by many investigations on natures of the doubly heavy tetraquark states and is strengthen again by the present work~\cite{potential,semay,valcarce}.

The states $\left[[bc]_{\bar{\mathbf{3}}_c}[\bar{u}\bar{d}]_{\mathbf{3}_c}\right]_{\mathbf{1}}$ with $00^+$ is allowed because of no symmetry restriction on the diquark $[bc]$. In the contrary to the diquarks $[bb]$ and $[cc]$, the color-magnetic interaction in the diquark $[bc]$ is weak attractive due to its spin $s_b=0$. Therefore, the energy of the state $\left[[bc]_{\bar{\mathbf{3}}_c}[\bar{u}\bar{d}]_{\mathbf{3}_c}\right]_{\mathbf{1}}$ with $00^+$ is 7 MeV lower than that of the state $\left[[bc]_{\bar{\mathbf{3}}_c}[\bar{u}\bar{d}]_{\mathbf{3}_c}\right]_{\mathbf{1}}$ with $01^+$. With respect to the state $\left[[bb]_{\bar{\mathbf{3}}_c}[\bar{u}\bar{d}]_{\mathbf{3}_c}\right]_{\mathbf{1}}$ with $01^-$, which involves one angular excitation allowed to occur between two $b$-quarks because of their large masses. Meanwhile, the diquark $[bb]$ has spin zero so that the color-magnetic interaction is weak attractive. Therefore, the state with $01^-$ has a lower mass than that of other states with negative parity.

In one word, there exists strong attractive interactions coming from the Coulomb interaction, the color-magnetic interaction and one Goldstone boson exchange (mainly $\pi$), which are more than 200 MeV, in the stable doubly heavy tetraquark states $\left[[QQ]_{\bar{\mathbf{3}}_c}[\bar{u}\bar{d}]_{\mathbf{3}_c}\right]_{\mathbf{1}}$ with $I=0$, see Table \ref{contributions}. Lattice QCD simulations on the doubly heavy tetraquark states $[QQ][\bar{u}\bar{d}]$ also indicated that the phase shifts in the isospin singlet channels suggest attractive interactions growing as $m_{\pi}$ decreases~\cite{lqcdheavy}. The state $\left[[bb]_{\bar{\mathbf{3}}_c}[\bar{u}\bar{s}]_{\mathbf{3}_c}\right]_{\mathbf{1}}$ with $\frac{1}{2}1^+$ is analogical to the state $\left[[bb]_{\bar{\mathbf{3}}_c}[\bar{u}\bar{d}]_{\mathbf{3}_c}\right]_{\mathbf{1}}$ with $01^+$ with the exception of the one Goldstone boson exchange ($K$ and $\eta$). The magnitude of the attractive interaction is weaker than the state $01^+$ because
of no $\pi$-meson exchange interaction in the state with $\frac{1}{2}1^+$.

In order to quantitatively understand the dynamical mechanism forming the stable doubly heavy tetraquark states, we calculate the contributions coming from the different piece of the Hamiltonian in the ACFTM to the binding energies of the stable states, which are presented in Table \ref{contributions}. One can find that the most of the binding energies come from meson exchange interactions, which are equal to the values in the tetraquark states. Once meson exchanges are switched off, some of stable states will vanish with the exception of the states $[bb][\bar{u}\bar{d}]$ with $01^+$ and $12^+$ and the state $[bb][\bar{u}\bar{s}]$ with $\frac{1}{2}1^+$, which become into weak bound states with binding energy of several and a dozen MeV. The reason is that the meson exchange between two light quarks in the states $[QQ][\bar{q}\bar{q}]$ does not occur in the threshold consisting of two $Q\bar{q}$ mesons. Therefore, the doubly heavy tetraquark states $[QQ][\bar{q}\bar{q}]$ provide an ideal field to research the interaction between the different quark interactions because the chiral symmetry is explicitly broken in the heavy sector but it is spontaneously broken in the light one.

The color-magnetic interaction also plays an important role in the formation of the stable doubly heavy tetraquark states with isopsin zero, which makes contributions to the binding energies ranging from 54 MeV to 208 MeV, see Table \ref{contributions}. The Coulomb interaction, independence of spin and isospin, universally produces extremely strong attractive interactions ranging from 550 MeV to 719 MeV in the stable doubly heavy tetraquark states, which can be understood by the small separations between any two particles $\langle\mathbf{r}_{ij}^2\rangle^{\frac{1}{2}}$, specially for the distance of two heavy quarks $\langle\mathbf{r}_{12}^2\rangle^{\frac{1}{2}}$, see Table \ref{rms}. However, the Coulomb interaction has no direct contribution to the binding energies of the heavy tetraquark states $[QQ][\bar{q}\bar{q}]$, see Table \ref{contributions}. It can be found that the contributions to the binding energies from the color-magnetic and Coulomb interactions amplify with the increase of the mass ratio $\frac{m_{Q}}{m_{q}}$ in the group of states $[QQ][\bar{q}\bar{q}]$ with the same $[\bar{q}\bar{q}]$ and $IJ^P$, such as the group $[cc][\bar{u}\bar{d}]$, $[bc][\bar{u}\bar{d}]$ and $[bb][\bar{u}\bar{d}]$ with $01^+$.

The states $[bc][\bar{u}\bar{d}]$ and $[bb][\bar{u}\bar{d}]$ with $12^+$ as stable states against strong interactions should be emphasized because of weak attractive meson exchange and repulsive color-magnetic interactions, which are different from the binding mechanism of the states with $I=0$. The kinetic energies make great contributions to the binding energies in the states with $12^+$, see Table \ref{contributions}. The reason is that the repulsive color-magnetic interaction and the motions of quarks prevent any two quarks from approaching each other. Meanwhile, the magnitude of the $\pi$-meson exchange in the states $[QQ][\bar{u}\bar{d}]$ with $12^+$ ($\langle\mathbf{\sigma}_i\cdot\mathbf{\sigma}_j\rangle$$\langle\mathbf{F}_i\cdot\mathbf{F}_j\rangle=1$) weaken to the $\frac{1}{9}$ of that in the states $[QQ][\bar{u}\bar{d}]$ with $01^+$ ($\langle\mathbf{\sigma}_i\cdot\mathbf{\sigma}_j\rangle$$\langle\mathbf{F}_i\cdot\mathbf{F}_j\rangle=9$). In this way, any two quarks should sit at far from each other, see Table \ref{rms}, so that the kinetic energies greatly reduce, about 400 MeV, comparing with those of the states with $01^+$. For the same reasons, the energies of other states $[QQ][\bar{q}\bar{q}]$ with $2^+$ in Table \ref{spectrum} is just a little higher than their corresponding threshold. With respect to the state $[bb][\bar{u}\bar{s}]$ with $\frac{1}{2}1^+$, one-Goldstone-boson-exchange interaction cannot provide large attraction because of lacking of $\pi$-meson exchange interaction. The system also reduces its kinetic energy to strengthen the stability of the system.

The quark model with twobody quadratic confinement potential by means of the Casimir scaling and other interactions in the ACFTM is directly extended to the heavy tetraquark states $[QQ][\bar{q}\bar{q}]$, see Table \ref{diff}, the energies are in general higher about 100 MeV than those given by the ACFTM because the Casimir scaling will lead to anticonfinement for some color structure in the multiquark system~\cite{anticonfinement}. Meanwhile, the model with twobody confinement potential is also known to be flawed phenomenologically because it leads to power law van der Waals forces between color-singlet hadrons, which will disappear automatically by taking account into the flip-flop potential in the LQCD simulation on the tetraquark states~\cite{flip-flop}. Comparing with the twobody confinement potential, the fourbody confinement potential based on the lattice picture push down the energy of the tetraquark states, about 100 MeV, and can provide the binding energies ranging from over 30 MeV to about 100 MeV, which is therefore universal dynamical mechanism forming stable doubly tetraquark states in the ACFTM. In other words, some states with binding energies below 100 MeV are not stable states anymore in the model with twobody confinement potential.

\section{summary}

We systematically study the doubly heavy tetraquark states $[QQ][\bar{q}\bar{q}]$ with diquark-antidiquark picture in order to search for all possible stable states against strong interactions in the ACFTM with a multibody confinement potential, $\sigma$-exchange, one-gluon-exchange and
one-Goldstone-boson-exchange interactions. The ACFTM model predicts that the tetraquark states $[cc][\bar{u}\bar{d}]$ with $01^+$, $[bc][\bar{u}\bar{d}]$ with $00^+$, $01^+$, and $12^+$, $[bb][\bar{u}\bar{d}]$ with $01^-$, $01^+$ and $12^+$, $[bb][\bar{q}'\bar{s}]$ with $\frac{1}{2}1^+$ are stable states against strong interactions. The tetraquark states $[bb][\bar{u}\bar{d}]$ with $01^+$ and $[bb][\bar{q}'\bar{s}]$ with $\frac{1}{2}1^+$ are the most promising stable doubly heavy tetraquark states, which should be explored in experiments in the near future. The strong decays of those stable doubly heavy tetraquark states are kinematically forbidden if they really exist. However, they can decay only weakly or electromagnetically and therefore they must have a small decay width.

The color configuration $\left[[QQ]_{\bar{\mathbf{3}}_c}[\bar{q}\bar{q}]_{\mathbf{3}_c}\right]_{\mathbf{1}}$ dominates the propertits of the doubly heavy tetraquark states, in which the diquark $[QQ]$ can be regarded as a basic building block because of their small sizes. The Coulomb interaction is very strong attractive and greatly reduce the energies of the doubly heavy tetraquark states $[QQ][\bar{q}\bar{q}]$. However, it has no direct contribution to the binding energies of the bound states. The multibody confinement potential based on the color flux-tube picture employs a collective degree of freedom whose dynamics play an important role in the formation of the bound states, which push down the energies of the doubly heavy tetraquark states about 100 MeV comparing the twobody one.

The doubly heavy tetraquark states $[QQ][\bar{u}\bar{d}]$ with $I=0$ are strong bound states and have the binding energies of the order of 100 MeV mainly coming from color-magnetic interaction and one-Goldstone-boson-exchange interaction. The doubly heavy tetraquark states $[QQ][\bar{u}\bar{d}]$ with $12^+$ are weak bound states because of weak meson exchange and repulsive color-magnetic interactions, which is formed mainly by reducing their kinetic energies. The strange state $[bb][\bar{u}\bar{s}]$ with $\frac{1}{2}1^+$ is similar to the state $[bb][\bar{u}\bar{d}]$ with $01^+$ and however a relative weak bound state because one-Goldstone-boson-exchange interaction cannot provide large attraction because of lacking of $\pi$-meson exchange interaction. The state also reduces its kinetic energy to strengthen the stability of the system.

Until now, none of stable doubly heavy states $[QQ][\bar{q}\bar{q}]$ has been observed in experiments and therefore more comprehensive investigations on their properties are still needed. The experimental detection and analysis of the doubly heavy tetraquark states will undoubtedly provide an invaluable opportunity to severely check the availability of the different theoretical models and, therefore, will allow one to makes more reliable theoretical predictions on the exotic hadronic spectra.

\acknowledgments
{This research is partly supported by the National Science Foundation of China under Contracts Nos. 11875226, 11775118, 11535005 and Fundamental Research Funds for the Central Universities under Contracts No. SWU118111.}


\begin{thebibliography}{99}
\bibitem{review} H.X. Chen, W. Chen, X. Liu and S.L. Zhu, Phys. Rep. \textbf{639}, 1 (2016);
 F.K. Guo, C. Hanhart, Ulf-G. Mei{\ss}ner, Q. Wang, Q. Zhao and B.S. Zou, Rev. Mod. Phys. \textbf{90}, 015004 (2018);
 S.L. Olsen, T. Skwarnicki and D. Zieminska, Rev. Mod. Phys. \textbf{90}, 015003 (2018).
\bibitem{charmed} C.R. Deng, J.L. Ping, H.X. Huang, and F. Wang, Phys. Rev. D \textbf{98}, 014026 (2018).
\bibitem{referee} M.N. Anwar, J. Ferretti, and E. Santopinto, Phys. Rev. D \textbf{98}, 094015 (2018).
\bibitem{potential} J.P. Ader, J.M. Richard, and P. Taxil, Phys. Rev. D \textbf{25}, 2370 (1982);
 J. L. Ballot and J. M. Richard, Phys. Lett. B \textbf{123}, 449 (1983).
\bibitem{bag} J. Carlson, L. Heller, and J.A. Tjon, Phys. Rev. D \textbf{37}, 744 (1988).
\bibitem{constituentmodel} J. Vijande, A. Valcarce, and K. Tsushima, Phys. Rev. D \textbf{74}, 054018 (2006);
 D.M. Brink and F.M. Stancu, Phys. Rev. D \textbf{57}, 6778 (1998);
 B. Silvestre-Brac and C. Semay, Z. Phys. C \textbf{59}, 457 (1993).
 S. Zouzou, B. Silvestre-Brac, C. Gignoux, and J. Richard, Z. Phys. C \textbf{30}, 457 (1986).
\bibitem{chiralperturbation} A.V. Manohar and M. B. Wise, Nucl. Phys. B \textbf{399}, 17 (1993).
\bibitem{string} J. Vijande, A. Valcarce, and J.-M. Richard, Phys. Rev. D \textbf{76}, 114013 (2007);
 C. Ay, J.M. Richard, and J.H. Rubinstein, Phys. Lett. B \textbf{674}, 227 (2009).
\bibitem{lqcdbb} P. Bicudo and M. Wagner, Phys. Rev. D \textbf{87}, 114511 (2013);
 Y. Ikeda et al, Phys. Lett. B \textbf{729}, 85 (2014).
\bibitem{qcdsum} M.L. Du, W. Chen, X.L. Chen and S.L. Zhu, Phys. Rev. D \textbf{87}, 014003 (2013);
 W. Chen, T.G. Steele, and S.L. Zhu, Phys. Rev. D \textbf{89}, 054037 (2014).
\bibitem{xicc} R. Aaij et al. (LHCb Collaboration), Phys. Rev. Lett. \textbf{119}, 112001 (2017).
\bibitem{karliner} M. Karliner and J. L. Rosner, Phys. Rev. Lett. 119, 202001 (2017).
\bibitem{Eichten} E.J. Eichten and C. Quigg, Phys. Rev. Lett. \textbf{119}, 202002 (2017).
\bibitem{Francis} A. Francis, R.J. Hudspith, R. Lewis, and K. Maltman, Phys. Rev. Lett. \textbf{118}, 142001 (2017).
\bibitem{bicudo} P. Bicudo, J. Scheunert, and M. Wagner, Phys. Rev. D \textbf{95}, 034502 (2017).
\bibitem{cqmpark} W. Park, S. Noh, and S.H. Lee, arXiv:1809.05257v1 [nucl-th].
\bibitem{unstable} A. Czarnecki, B. Leng, and M. B. Voloshin, Phys. Lett. B \textbf{778}, 233 (2018).
\bibitem{heavy} C. Quigg, arXiv:1804.04929 [hep-ph].
\bibitem{pseudoscalar} S.S. Agaev, K. Azizi, B. Barsbay, and H. Sundu, arXiv:1806.04447 [hep-ph].
\bibitem{decay} S.S. Agaev, K. Azizi, B. Barsbay, and H. Sundu, arXiv:1809.07791v1 [hep-ph].
\bibitem{francis} A. Francis, R.J. Hudspith, R. Lewis, and K. Maltman, arXiv:1810.10550 [hep-lat].
\bibitem{richard} J.-M. Richard, A. Valcarce, and J. Vijande, arXiv: 1811.02863 [hep-ph].
\bibitem{tcqm} N. Isgur and G. Karl, Phys. Rev. D \textbf{18}, 4187 (1978);
 Y. Fujiwara, C. Nakamoto, and Y. Suzuki, Phys. Rev. C \textbf{54}, 2180 (1996);
 Y. Fujiwara, K. Miyagawa, M. Kohno, Y. Suzuki, and C. Nakamoto, Nucl. Phys. A \textbf{737}, 243 (2004).
\bibitem{lqcd} C. Alexandrou, P. de Forcrand, and A. Tsapalis, Phys. Rev. D \textbf{65}, 054503 (2002);
 T.T. Takahashi, H. Suganuma, Y. Nemoto, and H. Matsufuru, Phys. Rev. D \textbf{65}, 114509 (2002);
 F. Okiharu, H. Suganuma, and T.T. Takahashi, Phys. Rev. Lett. \textbf{94}, 192001 (2005);
 A. Pimikov, H.J. Lee, N. Kochelev, and P.M. Zhang, Phys. Rev. D \textbf{95}, 071501(R) (2017).
\bibitem{ACFTM} J.L. Ping, C.R. Deng, F. Wang, and T. Goldman, Phys. Lett. B \textbf{659}, 607 (2008);
 C.R. Deng, J.L. Ping, F. Wang, and T. Goldman, Phys. Rev. D \textbf{82}, 074001 (2010);
 C.R. Deng, J.L. Ping, H. Wang, P. Zhou, and F. Wang, Phys. Rev. D \textbf{86}, 114035 (2012);
 C.R. Deng, J.L. Ping, Y.C. Yang, and F. Wang, Phys. Rev. D \textbf{86}, 014008 (2012);
 C.R. Deng, J.L. Ping, Y.C. Yang, and F. Wang, Phys. Rev. D \textbf{88}, 074007 (2013);
\bibitem{vandewaals} G. Feinberg and J. Sucher, Phys. Rev. D \textbf{20}, 1717 (1979);
 O.W. Greenberg and H.J. Lipkin, Nucl. Phys. A \textbf{370}, 349 (1981);
 J. Weinstein and N. Isgur, Phys. Rev. Lett. \textbf{48}, 659 (1982).
\bibitem{gauge} H.J. Lipkin, Phys. Lett. B 113, 490 (1982);
 O.W. Greenberg and J. Hietarinta, Phys. Lett. B 86, 309 (1979);
 D. Robson, Phys. Rev. D 35, 1018 (1987).
\bibitem{anticonfinement} V. Dmitrasinovic, Phys. Rev. D \textbf{67}, 114007 (2003).
\bibitem{SNChen} S.N. Chen, and J.L. Ping, Mod. Phys. Lett. A \textbf{27}, 1250025 (2012).
\bibitem{linear} J.L. Ping, F. Wang, and T. Goldman, Nucl. Phys. A \textbf{657}, 95 (1999);
 A. Valcarce, H. Garcilazo, F. Fern$\acute{a}$ndez, and P. Gonz$\acute{a}$lez, Rep. Prog. Phys. \textbf{68}, 965 (2005).
\bibitem{goldman} T. Goldman and S. Yankielowicz, Phys. Rev. D \textbf{12}, 2910 (1975).
\bibitem{kappa} G.S. Bali, Phys. Rev. D \textbf{62}, 114503 (2000);
  C. Semay, Eur. Phys. J. A \textbf{22}, 353 (2004);
  N. Cardoso, M. Cardoso, and P. Bicudo, Phys. Lett. B \textbf{710}, 343 (2012).
\bibitem{chnqm} A. Manohar and H. Georgi, Nucl. Phys. B \textbf{234}, 189 (1984).
\bibitem{su3} L.Y. Glozman and D.O. Riska, Phys. Rep. \textbf{268}, 263 (1996).
\bibitem{NNNY} Y. Fujiwara, C. Nakamoto, and Y. Suzuki, Phys. Rev. C \textbf{54}, 2180 (1996).
 Y. Fujiwara, T. Fujita, C. Nakamoto, Y. Suzuki, and M. Khono, Nucl. Phys. A \textbf{639}, 41c (1998);
 Y. Fujiwara, M. Kohno, C. Nakamoto, and Y. Suzuki, Phys. Rev. C \textbf{64}, 054001 (2001);
 Y. Fujiwara, T. Fujita, M. Kohno, C. Nakamoto, and Y. Suzuki, Phys. Rev. C \textbf{65}, 014002 (2001);
 Y. Fujiwara, K. Miyagawa, M. Khono, Y. Suzuki, and C. Nakamoto, Nucl. Phys. A \textbf{737}, 243 (2004).
\bibitem{chiralmeson} J. Vijande, F. Fernandez, and A. Valcarce, J. Phys. G \textbf{31}, 481 (2005).
\bibitem{weistein} J. Weinstein and N. Isgur, Phys. Rev. D \textbf {27}, 588 (1983).
\bibitem{masssigma} M.D. Scadron, Phys. Rev. D \textbf{26}, 239 (1982).
\bibitem{rel-meson} S. Godfrey and N. Isgur,  Phys. Rev. D \textbf{32}, 189 (1985).
\bibitem{three-models} C. Semay and B.Silvestre-Brac, Phys. Rev. D \textbf{46}, 5177 (1992) and therein.
\bibitem{qdcsm}  U. Straub, Z.Y. Zhang, K. Brauer, A. Faessler, and S.B. Khadkikar, Phys. Lett. B \textbf{200}, 241 (1988);
 F. Wang, G.H. Wu, L.J. Teng, and T. Goldman, Phys. Rev. Lett. \textbf{69}, 2901 (1992);
 G.H. Wu, L.J. Teng, J.L. Ping, F. Wang, and T. Goldman, Phys. Rev. C \textbf{53}, 1161 (1996);
 J.L. Ping, F. Wang, and T. Goldman, Nucl. Phys. A \textbf{657}, 95 (1999);
 J.L. Ping, H. R. Pang, F.Wang, and T. Goldman, Phys. Rev. C \textbf{65}, 044003 (2002).
 M. Chen, H.X. Huang, J.L. Ping, and F. Wang, Phys. Rev. C \textbf{83}, 015202 (2011);
 H.X. Huang, J.L. Ping, and F. Wang, Phys. Rev. C \textbf{89}, 034001 (2014).
 F. Huang and W.L. Wang, Phys. Rev. D \textbf{98}, 074018 (2018).
\bibitem{newhadrons} J. Vijande, E. Weissman, N. Barnea, and A. Valcarce, Phys. Rev. D \textbf{76}, 094022 (2007);
 J. Vijande, E. Weissman, A. Valcarce, and N. Barnea, Phys. Rev. D \textbf{76}, 094027 (2007);
 J. Vijande, A. Valcarce, and N. Barnea, Phys. Rev. D \textbf{79}, 074010 (2009).
 J.M. Richard, A. Valcarce, and J. Vijande, Phys. Lett. B \textbf{774}, 710 (2017);
 J.M. Richard, A. Valcarce, and J. Vijande, Phys. Lett. B \textbf{790}, 248 (2019).
\bibitem{GEM} E. Hiyama, Y. Kino, and M. Kamimura, Prog. Part. Nucl. Phys. \textbf{51} 223 (2003).
\bibitem{ebert} D. Ebert, R.N. Faustov, V.O. Galkin, and W. Lucha, Phys. Rev. D \textbf{76}, 114015 (2007).
\bibitem{ccqqvij} J. Vijande, F. Fern$\acute{a}$ndez, A. Valcarce, and B. Silvestre-Brac, Eur. Phys. J. A \textbf{19}, 383 (2004).
\bibitem{pepin} S. Pepin, Fl. Stancu, M. Genovese, and J.M. Richard, Phys. Lett. B \textbf{393}, 119 ( 1997).
\bibitem{sakai} S. Sakai, L. Roca, and E. Oset, Phys. Rev. D \textbf{96}, 054023 (2017).
\bibitem{qcdsumbcqq} W. Chen and T.G. Steele, arXiv:1310.8337 [hep-ph].
\bibitem{valcarce} J. Vijande, A. Valcarce, and N. Barnea, Phys. Rev. D \textbf{79}, 074010 (2009).
\bibitem{pflaumer} M. Pflaumer, P. Bicudob, M. Cardoso, A. Peters, and M. Wagner, arXiv:1811.04724v1 [hep-lat].
\bibitem{park} W. Park and S.H. Lee, Nucl. Phys. A \textbf{925}, 161, (2014).
\bibitem{semay} B. Silvestre-Brac and C. Semay, Z. Phys. C \textbf{57}, 273 (1993).
\bibitem{lqcdheavy} Y. Ikeda, B. Charron, S. Aoki, and T. Doi et al., Phys. Lett. B \textbf{729}, 85 (2014).
\bibitem{flip-flop} F. Okiharu, H. Suganuma, and T.T. Takahashi, Phys. Rev. D \textbf{72}, 014505 (2005);

\end{thebibliography}
\end{document}